\documentclass[11pt,twocolumn,letterpaper]{article}

\usepackage{cvpr}
\usepackage{times}
\usepackage{epsfig}
\usepackage{graphicx}
\usepackage{amsmath}
\usepackage{amssymb}
\usepackage[backend=biber]{biblatex}
\usepackage{subfig}
\usepackage{hyperref}
\hypersetup{
    colorlinks=true,
    linkcolor=blue,
    filecolor=magenta,      
    urlcolor=blue,
}
\graphicspath{{Images/}}
\usepackage{enumitem}
\pdfoutput=1

\cvprfinalcopy 

\def\cvprPaperID{****} 

\begin{document}

\title{Generative Models for Pose Transfer}

\author{Patrick Chao*\\
{\tt\small prc@berkeley.edu}\\
\and
Alexander Li*\\
{\tt\small alexli1@berkeley.edu}\\
\and
Gokul Swamy*\\
{\tt\small gokul.swamy@berkeley.edu}\\
}

\maketitle

\begin{abstract}
   We investigate nearest neighbor and generative models for transferring pose between persons. We take in a video of one person performing a sequence of actions and attempt to generate a video of another person performing the same actions. Our generative model (pix2pix) outperforms k-NN at both generating corresponding frames and generalizing outside the demonstrated action set. Our most salient contribution is determining a pipeline (pose detection, face detection, k-NN based pairing) that is effective at performing the desired task. We also detail several iterative improvements and failure modes.

\end{abstract}

\section{Introduction}
Pose transfer is a problem with a variety of applications that pushes the boundaries of image or video generation. First, being able to transfer poses is useful in situations where a person cannot perform a specific series of actions. This is especially valuable for video production and editing (\eg enhancing the footage produced by stunt doubles for a movie). The problem also is quite challenging from a computer vision perspective because of the difficulty of detecting joint locations reliably and being able to transfer pose from an arbitrary person to another. Lastly, pose transfer poses an ethical dilemma of whether we are able to trust the images and videos we see on the internet. The ``deepfakes" that have recently captured popular attention underscore the need for work in this area \cite{deep_fake}.

\subsection{Problem Description}
Our goal is to go from a video of individual $A$ performing a sequence of actions to a video of individual $B$ performing the same actions.

We developed three key qualitative metrics to measure the quality of our model output video:

\begin{enumerate}
    \item Smooth video: the individual frames will be calculated separately, so a high quality video will smoothly transition between frames. 
    \item Sharp edges: the outlines of a generated individual should be as sharp and realistic as possible
    \item Minimal artifacts: the model should avoid producing obvious visual artifacts that degrade the quality of the output
\end{enumerate}

We attempted to optimize our model to handle all three of these metrics as best as possible.

\subsection{Related Work}
Much of the previous computer vision research in human pose has focused on pose estimation, the process of extracting the joint locations from raw images \cite{deep_pose, cao2017realtime}. While pose estimation is a difficult task, these papers develop fairly robust models that can generate pose skeletons with relatively little noise. 

More recently, researchers have been working on learning-based approaches to problems relevant to pose transfer. One work focused on developing generative models for images in random poses, without regard to correspondence to some input sequence of poses \cite{deep_video_generation}. Another focuses on interpolating between different views of an object by generating dense correspondences between pixels \cite{view_morphing}. A final paper discussed using a two-stage process for synthesizing person images in arbitrary poses, generating a coarse structure before refining it \cite{pose_guided}.

We're interested in a learning-based model that can directly map poses from one individual to another, ideally avoiding the computational overhead of building and keeping track of a model of a subject. We examine in this paper the feasibility of direct generative approaches. 

\section{Methods}
In this section, we detail several key methods that act as building blocks for the pose transfer models that we tested.

\subsection{Pose Skeleton Generation}
\begin{figure}
\includegraphics[width=0.5\textwidth]{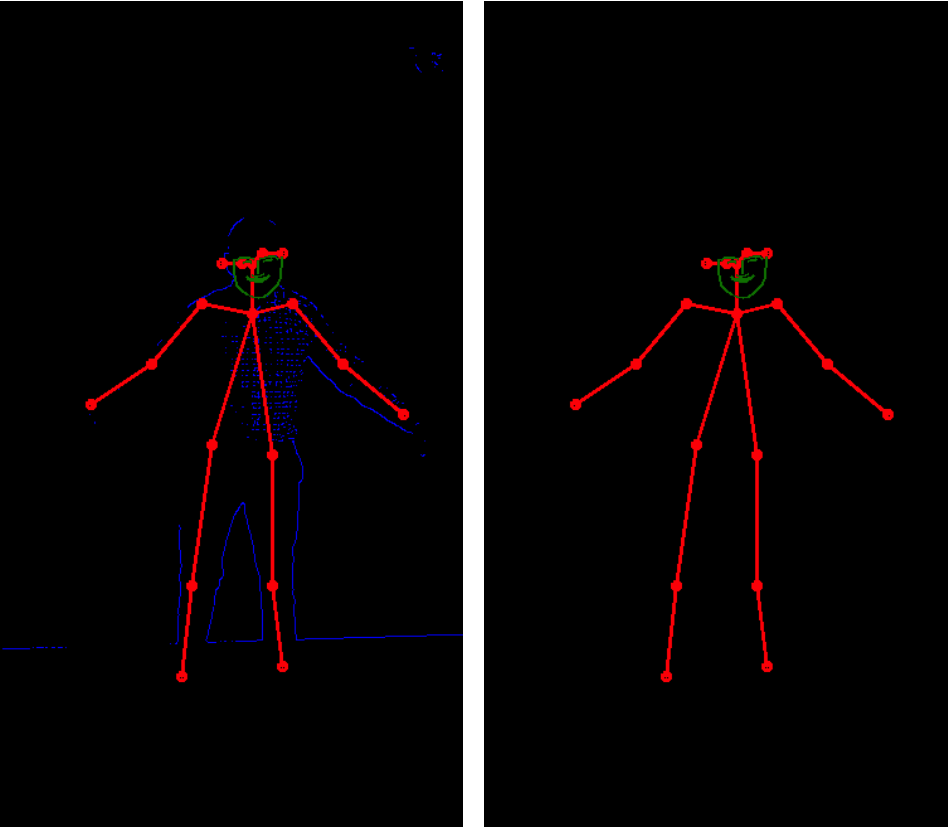}
\caption{Pose Skeletons. Left with edges, right without.}
\label{fig:pose skeletons}
\end{figure}
We use a Python implementation of a compressed version of OpenPose to generate our joint locations \cite{tf_pose_estimation}. This model provided the best trade-off in terms of accuracy and time during our testing process. We originally passed in just the joint locations but noticed that our model struggled with generating faces. To deal with this, we used a Python implementation of the standard dlib face detection library to add in contours of facial features \cite{face_recognition}. We used a HoG face detector when possible, opting for a more expensive CNN-based detector only when necessary. We also originally attempted to use the OpenCV implementation of Canny Edge Detection as an additional supervisory signal for training. We chose the upper and lower thresholds as $\frac{4}{3} \text{ * median}$ and $\frac{2}{3} \text{ * median}$, respectively. We combined each of these three outputs into a single image by placing each output on a different color channel, allowing the model to learn what inputs were important and which it could safely ignore. The output poses are demonstrated in Figure \ref{fig:pose skeletons} However, we noticed that including edges heavily reduced the ability of our model to take as input arbitrary pose skeletons due to overfitting on a particular person's outline. Pursuant to our original goal, we chose to not use edge detection in our final model. Overall, this process took around $1$ second per input frame.

\subsection{k-NN}
To calculate the similarity between poses of images, we quickly realized we needed to reduce the problem down to a pose representation of the image to calculate the intrinsic spatial relationship. Calculating the $L2$ distance between raw images would result in a very large number of computations that would scale according to the size of the image, which is also undesirable. This would also not accurately find similar poses across different training videos, due to differences in body shape, clothing, and background. 

Thus, we pivoted to generate pose matrices per video, a tensor containing the pose coordinates for each frame in a video. From the OpenPose library, we obtained a set of $18$ joints with $(x,y)$ coordinates, or an $18\times 2$ matrix per frame. This meant we had an $18\times 2\times \text{number of frames}$ tensor representing the joint coordinates for each frame. 

\begin{figure}
\begin{center}
\includegraphics[scale=0.30]{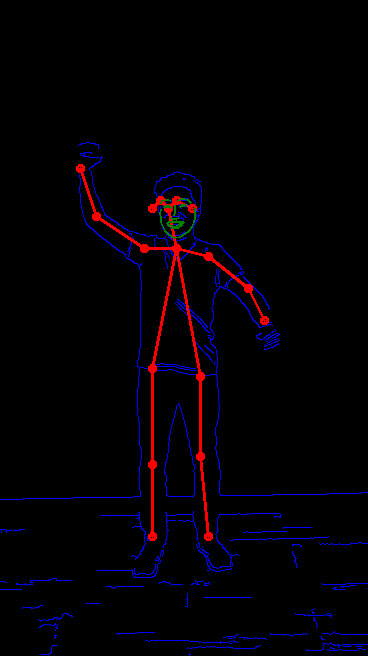} \includegraphics[scale=0.30]{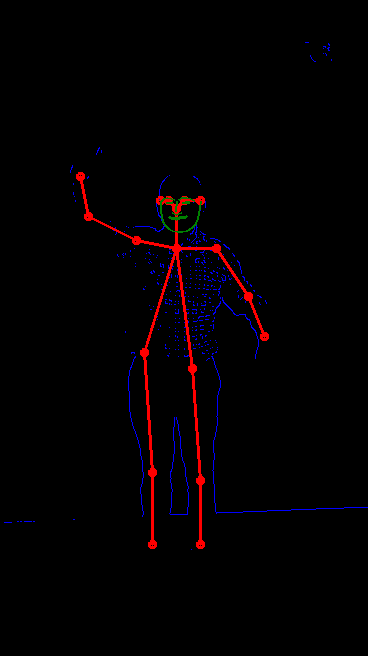}
\end{center}
\caption{Two Matched Poses through k-NN}
\label{fig:kNN Pose Match}
\end{figure}

From here, we utilized k-NN to determine the closest match of a given pose. From video $A$ and $B$, we generate two pose matrices $pose_A$ and $pose_B$. Given one pose matrix from $pose_A$, we search through all poses in $pose_B$ by taking the $\ell_2$ norm over each pose image and selecting the minimum $k$ entries. An example is seen in Figure \ref{fig:kNN Pose Match}.

We also ran into a few issues of missing joints. In some frames, individual joints are occluded such as when the individual crosses arms or turns his/her body, or body parts briefly go out of frame. This results in no joint being recorded. This would result in the given frames never being selected since they have joint positions unspecified at $(0,0)$. To prevent this, we replace all unseen joints with the median value for the given joint.

\subsection{Conditional Adversarial Networks}
Conditional adversarial networks (conditional GANs) solve the problem of generating outputs $y$ from a probability distribution that is conditioned on inputs $x$. Using a latent noise vector $z$, the generator learns the function $G: \{x, z\}\rightarrow y$. Meanwhile, the discriminator learns to distinguish images generated by the generator from natural images. For this paper, we used the pix2pix model \cite{pix2pix}, which uses several modifications to the standard conditional adversarial network to achieve realistic outputs. It uses a weighted sum of the standard GAN loss and the $L1$ loss, in order to discourage blurriness in the output:
\begin{equation}
    G^* = \arg \min_G \max_D \mathcal{L}_{cGAN}(G, D) + \lambda \mathcal{L}_{L1}(G)
\end{equation}

\begin{figure}
\begin{center}
\includegraphics[scale=0.95]{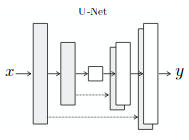}
\end{center}
\caption{Generator architecture for pix2pix}
\label{fig:pix2pix architecture}
\end{figure}

The pix2pix generator uses a ``U-net" architecture, as shown in Figure \ref{fig:pix2pix architecture}. It uses skip connections to preserve low level details, which are otherwise lost during the downsampling layers of the network.  

The discriminator does not evaluate the entire output image at once. Rather, it classifies each $N \times N$ patch in the image and averages the output. This is fairly accurate, and greatly reduces the number of parameters required. Overall, pix2pix achieves notably good results on a variety of image generation tasks, so we use it as the backbone for our pose transfer model. 

\section{Models}
We experimented with two major methods, a pose to picture model, and a picture to picture model. In the best case, we would like our model to take in an arbitrary video of an individual, and transfer it to another given individual. We first evaluated the feasibility of k-Nearest Neighbors as a baseline, before moving on to generative models for pose transfer. We developed pose2pics, where we generated the output image based on a lower dimensional pose skeleton. However, the pose space is slightly different for various individuals, and can lead to visual artifacts when predicting unseen images. To deal with these situations, we also experimented with training a direct pix2pix model end to end.

\subsection{Nearest Neighbor in Pose Space}
Our most basic model was just matching via k-NN on the locations of joints. Given enough training data of individuals in various positions, a k-NN model would perform perfectly in theory, being able to approximate any individual in any pose. However, we were interested in seeing the results in very limited training data environments where the individual may not explore much of the pose space.

For our experiments, we wanted to take a given video of $B$, and input video $A$ and generate an output video of individual $B$ in the same poses as $A$. For each frame of $A$, we selected the most similar frame of $B$. By concatenating these frames together, we obtain our k-NN predicted video. 

This approach produced several problems. First, individual frames in the output frequently differed drastically from one to another, even though the input frames have smooth transitions. This is due to noise in the output of the pose estimator. For example, if the previous $10$ pose frames in $A$ had all arm joints, but the next given frame was missing the left forearm joint, the closest neighbor output for the given frame may change dramatically, while the actual input video is very smooth.

On the other hand, if the input poses do not change drastically but there are two output poses that are similar on distance but different in terms of actual image, this would result in very jumpy transitions as well. This could occur if the poses for individual $B$ are not perfectly accurate or do not properly represent the image.

To prevent these defects, we introduced two major aspects: frame thresholding and motion interpolation.

\subsection{k-NN Frame Thresholding}
We utilized frame thresholding to reduce error from incorrect pose calculations in the second form of error mentioned above.

Consider frames $A_{t-1}$ and $A_{t}$, representing the previous and current frame from the input video we would like to match to. We would like to obtain $B_{t-1}$ and $B_{t}$ where these are the closest matches to $A_{t-1}$ and $A_t$. $B_{t-1}$ is our previous predicted output, and our $\hat B_t$ represents the predicted closest value to $A_t$ through k-NN. Without any thresholding, we would say that $B_t=\hat B_t$, and concatenate these frames. However, since small improvements in the $\ell_2$ pose distance may represent large changes in the output image, thus it is not always advantageous to select the image with the lowest value for $k$. Thus we only select $\hat B_t$ as $B_t$ if there is a noticeable improvement over $B_{t-1}$. 
\[
B_{t} = \left\{\begin{array}{ll}
        \hat B_{t}, &\text{if } d(A_t,\hat B_{t})<d(A_t,B_{t-1})-\lambda\\
        B_{t-1},  &\text{else} 
        \end{array}\right\} 
\]

 $\lambda$ represents how large the improvement is required to select a new frame. Thus larger values of $\lambda$ represent smoother frames but lower quality representations.

\subsection{k-NN Motion Interpolation}
We utilized motion interpolation to reduce error from incorrect pose calculations in the first form of error mentioned above.

By virtue of selecting the closest neighbor frame, the actual motion outputted will not be a smooth transition. Actual video frames involve very minute changes over time, but pose calculations move in discrete segments. To alleviate this problem, we introduced motion interpolation between frames to generate smoother transitions.

To do this, we utilized the Butterflow library \cite{butterflow}, which generates intermediate frames through optical flow and motion interpolation calculations. We found that this allowed us to utilize fewer intermediate frames in k-NN and interpolate the remainder, resulting in visually smoother transitions. However, we found that this motion interpolation is very slow computationally, requiring about $5$ seconds per frame. For real-time pose transfer, interpolating frequently is not feasible.

\subsection{pose2pics}
We determined that training a model to take as input a pose skeleton (facial features and joint locations) allows our model to be input-individual agnostic. 

We first run a pose and face detection algorithm on a frame of individual from a video to generate a ``pose skeleton." We then train pix2pix to invert this mapping (passing the skeleton as an input to the generator and training the discriminator with the original image). 

We found that from simply training on the poses of the output individual, the model had trouble extrapolating to poses of other individuals, as each individual has slightly different joint locations. Thus we instead trained our model to transform poses of an individual $A$ to images of individual $B$ in the corresponding position.

\begin{center}
    Pose to Image from $A$ to $B$
    \vspace{-5px}
\end{center}
\begin{enumerate}[itemsep=-1ex]
    \item Preprocess all the training videos for individuals $A$ and $B$ into pose matrices and pose frames
    \item Match similar poses for $A$ and $B$ through k-NN
    \item Train a pix2pix model using the pose of $A$ with the actual image for $B$ for images with similar poses
\end{enumerate}

The Pose2Pics method ran into several issues. Creating image pairs for conditional GAN training often does not perform well, as there may be missing joints in the skeleton generator. Furthermore, since we generate these pairs through k-NN, we can only train on poses that exist in both sample videos. If one of them has a lack of expression in the pose space, this Pose2Pics model will not be able to train on diverse and representative data. 

In addition, a skeleton is a greatly underdetermined system for an image, as it lacks any 3-D depth, shape of joints, and any information about fingers or face. This makes it difficult for the model to produce highly detailed outputs. 

These problems motivate the approach outlined in the next section.

\subsection{Direct pics2pics between Persons}
Real-time pose transfer is our ultimate goal. Thus, we explored using a conditional GAN model to directly translate the images of person A to images of person B in the corresponding position. This approach could potentially greatly speed up the transfer process, as it removes the need to run computationally expensive pose estimation models on the input image stream. Furthermore, it eliminates the impact of noise in the generated pose skeletons at test time, as we pass the full image as input. However, this model is not capable of going from one arbitrary person to another.

Thus an even simpler model is to just directly map input image to output image: given individual $A$ in a certain pose, output individual $B$ in the same pose. The pipeline is very similar to that of pose2pics. In order to create the image training pairs, we utilize the calculated poses to determine images with similar pose images, as simply calculating the $\ell_2$ distance between images is highly variant on other irrelevant aspects such as body shape and clothing color.

\begin{center}
    Image to Image from $A$ to $B$
    \vspace{-5px}
\end{center}
\begin{enumerate}[itemsep=-1ex]
    \item Preprocess all the training videos for individuals $A$ and $B$ into pose matrices and pose frames
    \item Match similar poses for $A$ and $B$ through k-NN
    \item Train a pix2pix model using the image of $A$ with the actual image for $B$ for images with similar poses
\end{enumerate}

For both generative approaches to pose transfer, we used the standard pix2pix architecture, training on a dataset of roughly $1000$ images for 200 epochs. 

\subsection{Aligning Outputs of Generative Models}
One issue we noticed with the outputs of our generative models that the output frames would each have a slight shift from the previous frame in terms of the location of the subject. We compensated for this by detecting the bounding box of the face of the subject in each image and aligning the images to the center of the bounding box. We calculated the location of the face in the first frame and then, for each later frame, cropped the image until the face centers were at the same location and padded the image with copies of the last row or column.

\section{Results}
A folder of our output gifs and videos can be found \href{https://tinyurl.com/280posevideos}{here}. 

\subsection{Nearest Neighbors}

\begin{figure}
\begin{center}
\includegraphics[scale=0.28]{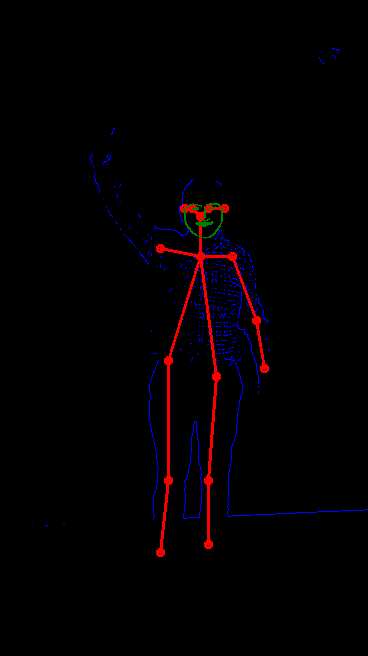} \includegraphics[scale=0.28]{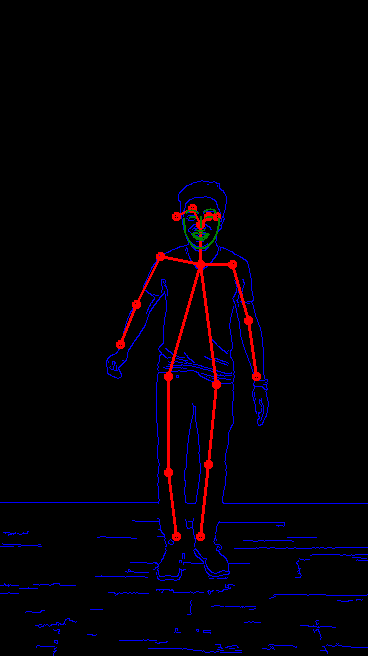}\\
\includegraphics[scale=0.28]{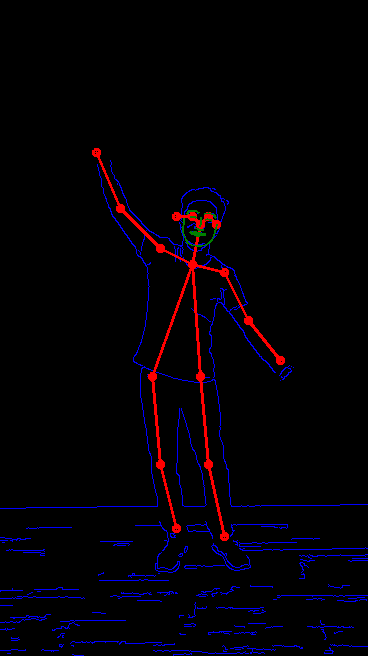}
\includegraphics[scale=0.28]{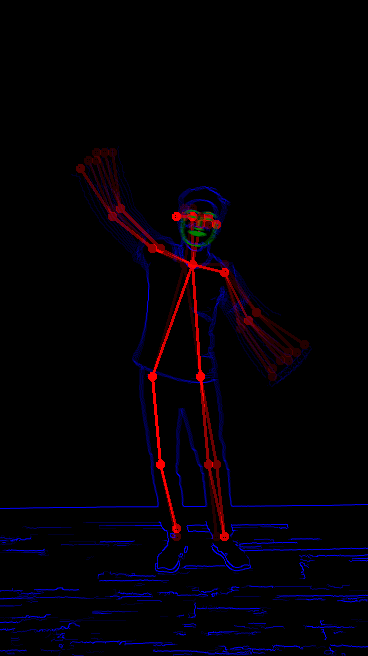}
\end{center}
\caption{Top Left: Pose to be Matched, Top Right: 1-NN without thresholding, Bottom Left: 1-NN with thresholding, Bottom Right: 7-NN with thresholding}
\label{fig:kNN joint missing}
\end{figure}

With k-NN, we found that mapping every input frame to an output frame resulted in highly jumpy videos, since each frame could result in a greatly varying closest neighbor. As stated in Section 3.3, this is because of the noise in the pose skeleton generator. 

We also ran into issues where the pose generator was unable to recognize joints in the extremities, resulting in gaps for pose skeletons for some frames. This resulted in a very low quality output, as k-NN is missing crucial information it needs to know which image is ``best" in the pose space. We attempted to only utilize frames where all the joints were present, but this resulted in large sections of continuous frames being removed. In Figure \ref{fig:kNN joint missing}, the top left image is the pose we would like to match, but its pose skeleton fails to capture the right arm raised. This results in a very poor matching, as shown by the large difference in the arm position between the top left and top right images. However, with our thresholding technique, we continue to utilize the previous outputs. These are far more accurate, since the input frames are temporally correlated. The bottom row of Figure \ref{fig:kNN joint missing} shows the resulting images for $k=1$ and $k=7$.

\subsection{pose2pics}

\begin{figure}
\begin{center}
\includegraphics[width=0.32\columnwidth, height =0.5\columnwidth]{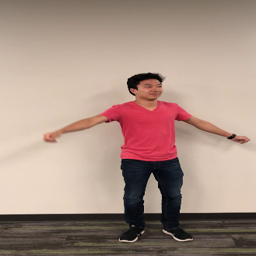}
\includegraphics[width=0.32\columnwidth, height =0.5\columnwidth]{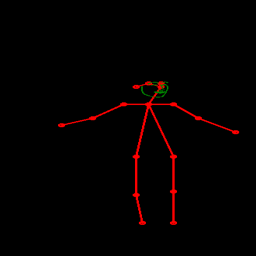}
\includegraphics[width=0.32\columnwidth, height =0.5\columnwidth]{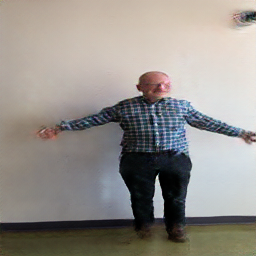}\\
\includegraphics[width=0.32\columnwidth, height =0.5\columnwidth]{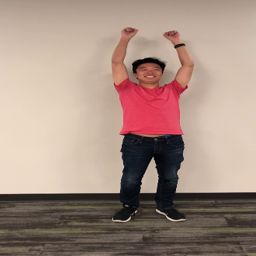}
\includegraphics[width=0.32\columnwidth, height =0.5\columnwidth]{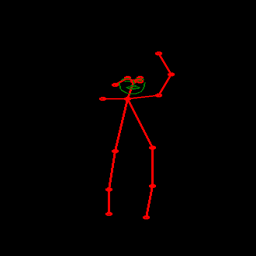}
\includegraphics[width=0.32\columnwidth, height =0.5\columnwidth]{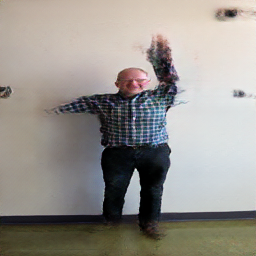}
\end{center}
\caption{Results for pose2pics. Left: input images, Middle: corresponding pose skeleton, Right: generated images}
\label{fig:Pose2pics}
\end{figure}

The pose2pics model sacrificed some accuracy and information in the input to generalize the input space. In Figure \ref{fig:Pose2pics}, the first row demonstrates an effective transfer, where the pose model accurately represents the true pose, and it translates accordingly. Professor Alexei Efros is in the same pose, and the output only contains minor artifacts. However, this model is also more susceptible to unforeseen situations such as loss of limbs in the pose skeleton. In the second row, we see that the right arm of the individual was not captured by the pose skeleton, making the pose2pics model truncate the arm at the shoulder in the generated image. This first demonstrates that the model heavily relies on the joint spatial data as a framework to generate the true image, and secondly demonstrates that improper training/testing data results in greatly skewed outputs. We also noticed that when we passed in skeletons generated from someone with a significantly different skeletal structure (for example, someone who was a foot taller than the person whose skeletons we originally trained on), the output image would have a clear physical abnormality (for example, elongated limbs). To deal with this, we could either train on a more diverse pose skeleton set (hopefully inducing the model to learn limb-length-invariant transformations) or train the model on joint angles instead of locations. We also noticed that pose2pics struggled with generalizing to never before seen positions. A more diverse training set might have helped mitigate this problem.

\subsection{pics2pics}

\begin{figure}
\begin{center}
\includegraphics[width=0.37\columnwidth, height =0.5\columnwidth]{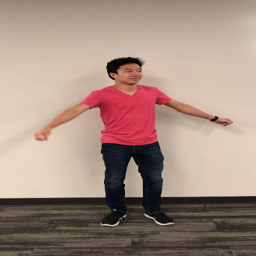}
\includegraphics[width=0.37\columnwidth, height =0.5\columnwidth]{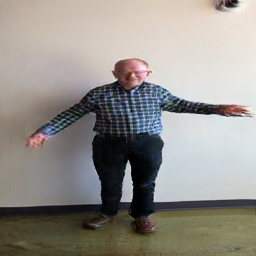} \\
\includegraphics[width=0.37\columnwidth, height =0.5\columnwidth]{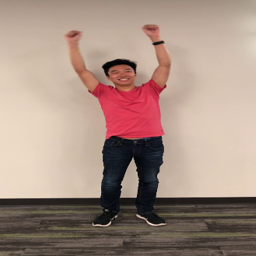}
\includegraphics[width=0.37\columnwidth, height =0.5\columnwidth]{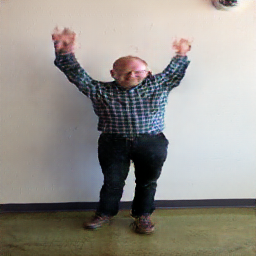}
\end{center}
\caption{Results for the direct pix2pix. The left column contains the input images, and the right contains the generated results. The top row shows an example of a good output result, while the bottom shows a deformed output, corresponding to a pose that is fairly different from any images seen in the training video of Professor Efros.}
\label{fig:direct pix2pix}
\end{figure}

With direct pics2pics, we found that we obtained quite promising results in terms of images. They seem to be quite realistic and have sharp edges, as seen in Figure \ref{fig:direct pix2pix}. In large part, this is due to its avoidance of the pose skeleton generator as an intermediate step, which introduces noise and greatly reduces signal. We also observe generalization in the conditional GAN, which has learned to raise both of Efros's arms above his head, even though it has never seen him do so in the training data. 

However, this method is not as invariant to background or clothing as other methods are, since it depends on a very strict input format of a particular individual in a given position. This model would be good for simple applications where there is lengthy and detailed training video of an individual, and the model solely serves the purpose of transferring poses between two specific people. This could be useful for producing images of an actor performing a trick from a video of a stunt double performing the same actions, as mentioned in our introduction.

\section{Conclusions and Future Work}
In this paper, we explored the application of the k-NN, pose2pics, and pics2pics models to the problem of pose transfer between videos. 

As a baseline, k-NN identified the pose skeleton generator as the weak link in our pose transfer pipeline. Noisiness in the pose skeletons was responsible for the jumpy transitions and vastly different input-output matchings. This made k-NN pose transfer difficult, even for sections of the pose space where our training data was concentrated. 

However, we were still able to extricate some useful lessons and promising beginnings from our experimental results. The generated images for the pose2pics model demonstrated that it was possible for the neural network to learn the structure of a person, which it could then use to generate an image when given a particular pose skeleton. Furthermore, the pics2pics model proved that the conditional GAN has the ability to generalize from training data to generate images that are out of the training distribution. We anticipate that a perfect pose skeleton generation algorithm would be able to generate better training pairs, further increasing the efficacy of both models. 

One obvious area of improvement would be reducing the time-complexity of our methods so we could generate videos in around real time. We implemented a webcam-based version of the previously discussed pose2pix pipeline but were not able to reduce frame generation time below one second, 30 times slower than the typical frame rate.
\begin{figure}
\includegraphics[width=\columnwidth]{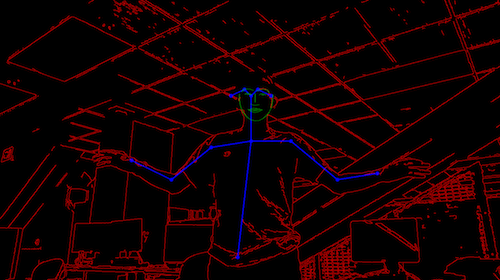}
\caption{Sample Output of Webcam-Based Pose Detection}
\end{figure}

One particularly exciting approach to explore for pose transfer between videos is to take advantage of temporal consistency between frames. Between one instant and the next, the video is unlikely to change much (given that we record at 60 FPS). We can maintain an underlying structure behind the frames. Then, using compressed neural-network based approaches or even purely algorithmic approaches (like a Kalman Filter), we can generate the difference between the current frame and the next. This way, we would be able to produce smooth videos while only having to perform computation at the boundaries of moving objects. Ideally, this would move us closer to the goal of real-time transfer. 

Additionally, while we chose to not include edge detection in our pipeline, one could produce more realistic outputs if they were able to find the right balance between enhanced supervision and the overfitting that edge detection provides. A potential way to do this would be to pass in pose skeletons from a wide variety of input sources so that the model learns to ignore the non-crucial parts of the edge image.

A final improvement would be to use a better pose estimation model. We did most of the project using standard tools such as OpenPose, but noise in the pose outputs caused most of the problems that affected our k-NN, Pose2Pics, and Pics2Pics models. We recently found a more powerful model \cite{VNect} that is much more accurate. It generates 3D pose estimates from 2D video, and also uses the 1 euro filter to avoid the problem of noisy pose estimates. With this pose estimator, we can design a more robust k-NN method that utilizes distances in the joint angle space, rather than the joint position space. If we try translating between individuals of vastly different heights --- such as Professor Joshua Hug to Professor John DeNero --- k-NN in the joint position space simply cannot accurately pair images. By working in the angle space, our pose transfer method can be invariant to differences in physical attributes. This stronger k-NN would generate much better training pairs, so the conditional GAN can train on perfect data.

\printbibliography

\end{document}